\documentclass[aps,prd,twocolumn,superscriptaddress,preprintnumbers,floatfix,nofootinbib,notitlepage,showkeys]{revtex4-1}

\usepackage[utf8]{inputenc}
\usepackage{cancel}

\usepackage{graphicx}
\usepackage{hyperref}
\usepackage{latexsym}
\usepackage{amsmath}
\usepackage{amssymb}
\usepackage{bbm}

\usepackage{ulem}
\usepackage{pdfsync}
\usepackage{epsfig}
\usepackage{epstopdf}
\usepackage{subfigure}
\usepackage{color}
\usepackage{comment}
\usepackage{slashed}

\def\beq{\begin{equation}}
\def\eeq{\end{equation}}
\def\baq{\begin{eqnarray}}
\def\eaq{\end{eqnarray}}

\newcommand{\be}{\begin{equation}} 
\newcommand{\ee}{\end{equation}}
\newcommand{\bea}{\begin{eqnarray}} 
\newcommand{\eea}{\end{eqnarray}}

\newcommand{\bmp}{\noindent\begin{minipage}{16cm}}
\newcommand{\emp}{\end{minipage}\vskip 7mm} 
\def\lsim{\mathrel{\raise.3ex\hbox{$<$\kern-.75em\lower1ex\hbox{$\sim$}}}}
\def\gsim{\mathrel{\raise.3ex\hbox{$>$\kern-.75em\lower1ex\hbox{$\sim$}}}}

\newcommand{\intron}[1]{}

\def\MP{M_{\rm P}}

\begin{document}

\title{The Palatini side of inflationary attractors}

\author{Laur J\"arv}
\email{laur.jarv@ut.ee}
\affiliation{Laboratory of Theoretical Physics, Institute of Physics,
University of Tartu,\\ W.\ Ostwaldi 1, 50411 Tartu, Estonia}
\author{Antonio Racioppi}
\email{antonio.racioppi@kbfi.ee}
\affiliation{National Institute of Chemical Physics and Biophysics, \\
						R\"avala 10, 10143 Tallinn, Estonia}
\author{Tommi Tenkanen}
\email{t.tenkanen@qmul.ac.uk}
\affiliation{Astronomy Unit, Queen Mary University of London, \\
                      Mile End Road, London, E1 4NS, United Kingdom}

\begin{abstract}
We perform an analysis of models of chaotic inflation where the inflaton field $\phi$ is coupled non-minimally to gravity via $\xi \phi^n g^{\mu\nu}R_{\mu\nu}(\Gamma), n>0$. We focus on the {\it Palatini} theory of gravity, i.e. the case where the assumptions of the General Relativity are relaxed (that of the connection being the Levi-Civita one) and the gravitational degrees of freedom are encoded not only in the metric but also the connection $\Gamma$, which is treated as an independent variable. We show that in this case the famous attractor behaviour of simple non-minimally coupled models of inflation is lost. Therefore the attractors are not universal 
but their existence depends on the underlying theory of gravity in a subtle way. We discuss what this means for chaotic models and their observational consequences.

\end{abstract}

%
\maketitle

%
\section{Introduction}

Cosmic inflation is the current paradigm for explaining the origin of the small inhomogeneities observed in the Cosmic Microwave Background (CMB) radiation~\cite{Starobinsky:1980te, Sato:1980yn, Guth:1980zm, Linde:1981mu, Albrecht:1982wi, Linde:1983gd,Lyth:1998xn,Mazumdar:2010sa,Martin:2013tda,Patrignani:2016xqp}. Amongst the parameters that are relevant to inflationary perturbations, two have been measured: the amplitude of the curvature power spectrum, $A_s=(2.141\pm 0.052)\times 10^{-9}$, and the corresponding spectral tilt, $n_s=0.9681\pm 0.0044$ (both at $1\sigma$ level) \cite{Ade:2015xua}. On top of that, the most recent analysis of BICEP2/Keck Array data has given an upper bound for the tensor-to-scalar ratio $r<0.09$ (at $2\sigma$ level) \cite{Array:2015xqh}. These are numbers that any successful model of inflation has to predict.

In this paper we investigate models of inflation with a non-minimal coupling to gravity of the type $\xi_n \phi^n g^{\mu\nu}R_{\mu\nu}$, where $\phi$ is the inflaton field, $g_{\mu\nu}$ is the space-time metric, $R_{\mu\nu}$ the Ricci tensor, $\xi_n$ a coupling constant, and $n>0$. Similar models have been discussed in a large number of works over the past decades, for example in \cite{Futamase:1987ua,Salopek:1988qh,Fakir:1990eg,Amendola:1990nn,Kaiser:1994vs,Bezrukov:2007ep,Bauer:2008zj,Park:2008hz,Linde:2011nh,Kaiser:2013sna,Kallosh:2013maa,Kallosh:2013daa,Kallosh:2013tua,Kallosh:2013tua,Galante:2014ifa,Chiba:2014sva,Boubekeur:2015xza,Pieroni:2015cma,Jarv:2016sow,Salvio:2017xul,Rasanen:2017ivk,Tenkanen:2017jih,Racioppi:2017spw}. These models are of particular interest, as non-minimal couplings should be seen as a generic ingredient of coherent model frameworks, generated by quantum corrections in a curved space-time \cite{Birrell:1982ix}. In particular, this is the case for the scenario where the Standard Model (SM) Higgs is the inflaton field \cite{Bezrukov:2007ep}.

Comparisons of non-minimally coupled chaotic models of inflation were performed in e.g.\ \cite{Linde:2011nh,Kaiser:2013sna,Kallosh:2013maa,Kallosh:2013daa,Kallosh:2013tua,Galante:2014ifa,Jarv:2016sow}. In Refs. \cite{Kaiser:2013sna,Kallosh:2013tua}, it was found that for large values of the non-minimal coupling strength all models, independently of the original scalar potential, asymptote to a universal attractor: the Starobinsky model \cite{Starobinsky:1980te}. Such a model can be approximated in terms of a canonically normalized scalar field in the Einstein frame with an exponentially flat potential
\be
\label{eq:starobinsky_nsr}
{\cal L}_E \simeq \sqrt{-g} \left[ -\frac{1}{2} M_{\rm P}^2 R+ \frac{1}{2} (\partial \chi )^2-  
M^4 \left(1-e^{-\sqrt{2\over 3} \frac{\chi}{M_{\rm P}}} \right) \right] ,
\ee
where $M_{\rm P}$ and $M$ are, respectively, the reduced Planck mass and a normalization scale for the scalar potential to be fitted to the observed amplitude of the curvature power spectrum.
The model can be shown to give the following predictions for inflationary observables
\bea
\label{starobinsky_nsr}
n_s &=& 1-\frac{2}{N}-\frac{9}{2N^2}, \\ \nonumber
r &=& \frac{12}{N^2},
\eea
where $N$ is the number of e-folds. We see that for $N=60$ the models predict $n_s=0.965, r=0.0033$. 
Simple non-minimally coupled chaotic inflation models therefore not only exhibit attractor-type universality but are also in perfect agreement with the Planck and BICEP2/Keck Array results.

However, as we will show, this 
applies only in the so-called metric formulation of gravity, and not in more general theories. In this paper we will demonstrate this in the following way: instead of assuming the usual metric case, where the connection is the Levi-Civita one, we allow for the connection to be an independent variable, i.e. study the dynamics of inflation in the context of {\it Palatini gravity}. Even though the metric and Palatini theories coincide within the General Theory of Relativity, in more general models these two formalisms lead to two inherently different theories of gravity \cite{Flanagan:2003rb,Olmo:2004hj,Olmo:2005,Allemandi2006,Sotiriou:2006hs,Fay:2007gg,Uddin:2007gj,Iglesias:2007nv,Tsujikawa:2007tg,Sotiriou:2008rp,DeFelice:2010aj,Vitagliano:2010pq,Tamanini:2010uq,Bauer:2010jg,Olmo:2011uz,Capozziello:2011et}. In particular, this means that models of inflation with non-minimal couplings to gravity cannot be characterized just by the form of the inflaton potential, but one needs to specify also the fundamental gravitational degrees of freedom, as was pointed out in e.g. \cite{Bauer:2008zj} and recently studied in more detail in \cite{Rasanen:2017ivk,Tenkanen:2017jih,Racioppi:2017spw,Markkanen:2017tun}.


\section{Non-minimally coupled models of inflation}
\label{inflation}

We consider a general class of inflationary models coupled non-minimally to Palatini gravity\footnote{For the purpose of this article we will keep the discussion of Palatini gravity and its differences to the metric case as minimal as possible. For the reader interested in further details on the topic, we suggest \cite{Bauer:2008zj} and references therein.} via
\be
\label{nonminimal_action}
S_J \! = \!\! \int \!\! d^4x \sqrt{-g}\left[\frac{g^{\mu\nu}}{2} \bigg(\partial_{\mu}\phi\partial_{\nu}\phi - f(\phi) R_{\mu\nu}(\Gamma)\bigg) - V(\phi)\right] ,
\ee
where $R_{\mu\nu}$ is the Ricci tensor, $f(\phi)>0$ is a non-minimal coupling function, $g_{\mu\nu}$ is the metric tensor and $g$ its determinant, and $\Gamma$ is the connection. 
  Working in the Jordan frame, the difference between the two formalisms arises in a different definition of $\Gamma$. In the Palatini formalism both $g_{\mu\nu}$ and $\Gamma$ are treated as independent variables, and the only assumption is that the connection is torsion-free, $\Gamma^\lambda_{\alpha\beta}=\Gamma^\lambda_{\beta\alpha}$. In this case the equations of motion will set the connection as a function of the metric and the scalar field: ${\Gamma}={\Gamma}(g^{\mu\nu},\phi)$ (see e.g. \cite{Bauer:2008zj}). This is in contrast to the metric formulation, where the connection is determined uniquely as a function of just the metric tensor, i.e.\ it is the Levi-Civita connection ${\Gamma}=\bar{\Gamma}(g^{\mu\nu})$. However,
 it is useful to perform a transformation to a frame where the non-minimal couplings in the Jordan frame action \eqref{nonminimal_action} vanish and the equations of motion for $\phi$ and $R_{\mu\nu}$ decouple. This can be done by performing a conformal transformation to the the so-called Einstein frame
\begin{equation}
\label{Omega}
g_{\mu\nu} = \Omega^{-1}(\phi)\tilde{g}_{\mu\nu}, \hspace{.5cm} \Omega(\phi)\equiv \frac{f(\phi)}{M_{\rm P}^2} \, ,
\end{equation}	
and the resulting field operator can be brought into a canonically normalized form by redefining the field operator by
\be
\label{chi}
\frac{d \chi}{d \phi} = \sqrt{\frac{1}{\Omega(\phi)}}\, .
\ee

As a result of these transformations, the action \eqref{nonminimal_action} becomes
\be
S_{\rm E} = \int d^4x \sqrt{-\tilde{g}}\bigg(\frac{1}{2}{\partial}_{\mu}\chi{\partial}^{\mu}\chi -\frac{1}{2}M_{\rm P}^2 \tilde{R} - U(\chi)  \bigg),
\label{EframeS}
\ee
where
\be
U(\chi) = \frac{V(\phi(\chi))}{\Omega^{2}(\phi(\chi))},
\label{eq:U}
\ee
is the Einstein frame potential and $\tilde{R} = \tilde{g}^{\mu\nu}\tilde{R}_{\mu\nu}(\bar{\Gamma})$, i.e. in the Einstein frame we retain the standard Levi-Civita connection. The procedure is similar in the metric gravity: the only difference is a more complicated form for the field redefinition \eqref{chi}
\be
\label{chi_metric}
\left. \frac{d \chi}{d \phi} \right|_\text{metric} = \sqrt{\frac{3}{2}\left(\frac{M_P}{\Omega}\frac{d \Omega}{d\phi}\right)^2+\frac{1}{\Omega}}\,,
\ee
 which eventually leads to a different potential for the canonically normalized field $\chi$, as we will show. In the Einstein frame, the difference between the two formulations is moved into the $\chi$ sector (see e.g. \cite{Bauer:2008zj}).

Assuming slow-roll, the inflationary dynamics is characterized by the usual slow-roll parameters and the total number of e-folds during inflation\footnote{The exact number of required e-folds is model-dependent and depends on the reheating mechanism. Here we concentrate only on the dynamics during inflation but the treatment is straightforward to augment with an analysis of reheating in any given model which specifies the coupling(s) between the inflaton and SM particles.}. The slow-roll parameters are defined in terms of the Einstein frame potential by
\be
\epsilon \equiv \frac{1}{2}M_{\rm P}^2 \left(\frac{1}{U}\frac{{\rm d}U}{{\rm d}\chi}\right)^2 \,, \quad
\eta \equiv M_{\rm P}^2 \frac{1}{U}\frac{{\rm d}^2U}{{\rm d}\chi^2} \,,
\ee
and the number of e-folds by
\be
N = \frac{1}{M_{\rm P}^2} \int_{\chi_f}^{\chi_i} {\rm d}\chi \, U \left(\frac{{\rm d}U}{{\rm d} \chi}\right)^{-1},
\label{Ndef}
\ee
where the field value at the end of inflation, $\chi_f$, is defined via $\epsilon(\chi_f) = 1$.  
The field value $\chi_i$ at the time a given scale left the horizon is given by the corresponding $N$. 

To obtain the correct amplitude for the curvature power spectrum, the potential has to satisfy \cite{Lyth:1998xn,Ade:2015xua}
\be
\label{cobe}
\frac{U(\chi_i)}{\epsilon(\chi_i)} = (0.027M_{\rm P})^4 ,
\ee
and the other two main observables, i.e. the spectral index and the tensor-to-scalar ratio are given in terms of the slow-roll parameters by
\bea
n_s &\simeq& 1+2\eta-6\epsilon \\ \nonumber
r &\simeq& 16\epsilon ,
\eea
respectively.


Let us now compare the predictions for observables in metric and Palatini theories in a particular class of non-minimally coupled chaotic inflation models. The Jordan frame actions that we consider\footnote{In this article we ignore the role of quantum corrections, as they are assumed to be subdominant. For some literature on the topic see for instance \cite{Kannike:2015apa,Kannike:2015kda,Marzola:2015xbh,Marzola:2016xgb,Artymowski:2016dlz,
Racioppi:2017spw,Markkanen:2017tun,Racioppi:2018zoy} and references therein.} are characterized by
\begin{eqnarray}
\label{eq:f} 
f(\phi) &=& \MP^2 \left[ 1 + \xi  \left( \frac{\phi}{\MP} \right)^n \right] \, ,\\
V(\phi) &=& \lambda_{2n} M_*^{4-2n} \phi^{2n} \label{eq:V}
\end{eqnarray}
where $n>0$, $\lambda_{2n}$ is a dimensionless coupling and $M_*^{4-2n}$ is a dimensionful term needed to have a scalar potential with mass dimension equal to four. The constraint on the amplitude of scalar perturbations (\ref{cobe})
allows us to fix only the product $\lambda_{2n} \, M_*^{4-2n}$. Therefore, it is customary to take $M_*=M_P$ and then fix $\lambda_{2n}$ to an explicit numerical value \cite{Ade:2015lrj}. Furthermore, from \eqref{eq:f} we also see that when $\phi\to 0$ after inflation and subsequent reheating, one retains the pure GR form of the theory.

It is well known that in metric gravity, the model defined by Eqs. (\ref{eq:f}) and (\ref{eq:V}), gives for large $\phi$ the following Einstein frame potential \cite{Kallosh:2013tua}
\begin{equation}
\label{eq:metric:attractor}
U(\chi)_{\rm metric} \simeq \frac{\lambda_{2n} \, M_P^{4}}{\xi^2} \left(1-e^{-2\sqrt{\frac{\xi}{1+6\xi}} \frac{\chi}{M_P}} \right) ,
\end{equation}
and therefore predicts Starobinsky inflation (see Eq. (\ref{eq:starobinsky_nsr})) for $\xi \to \infty$. Apart from the prefactor $\lambda_{2n}$, any dependence on $n$ cancels out because of the large field limit and the corresponding field redefinition in metric gravity \cite{Kallosh:2013tua}. 
We also see that the potential (\ref{eq:metric:attractor}) is equivalent to the well-known $\alpha$-attractor models \cite{Galante:2014ifa} for
\begin{equation}
\label{eq:alpha}
\alpha=1 + \frac{1}{6\xi} .
\end{equation}

On the other hand, in Palatini gravity the scalar field redefinition \eqref{chi} yields after integration
\begin{equation}
\label{eq:hypergeom}
\chi= \phi F \left(\tfrac{1}{2}, \tfrac{1}{n}; \tfrac{n+1}{n},-\xi \left(\tfrac{\phi}{\MP}\right)^n \right) \,,
\end{equation}
where $F$ is a hypergeometric function. For general $n$ it is difficult to find a nice analytic expression for the potential $U(\phi(\chi))$, but for e.g.\ $n=2$ we recover \cite{Bauer:2008zj}
\begin{equation}
U(\chi) \simeq \frac{\lambda_4 \MP^4}{\xi^2} \left(1-8 e^{-\frac{2 \sqrt{\xi} \chi}{\MP}} \right) \,,
\label{eq:Palatini:attractor:4}
\end{equation}
while $n=1$ yields
\begin{equation}
U(\chi) \simeq \frac{\lambda_2 \MP^4}{\xi^2} \left(1- \frac{8 \MP^2}{(2 \MP + \xi \chi)^2}\right) \,.
\end{equation}
Note the different $\xi$ dependence in the exponentials in Eqs. (\ref{eq:metric:attractor}) and (\ref{eq:Palatini:attractor:4}).
Analogous power laws appear also for other values of $n$.

\begin{figure}[t!]
  \begin{center}
\includegraphics[width=0.45 \textwidth]{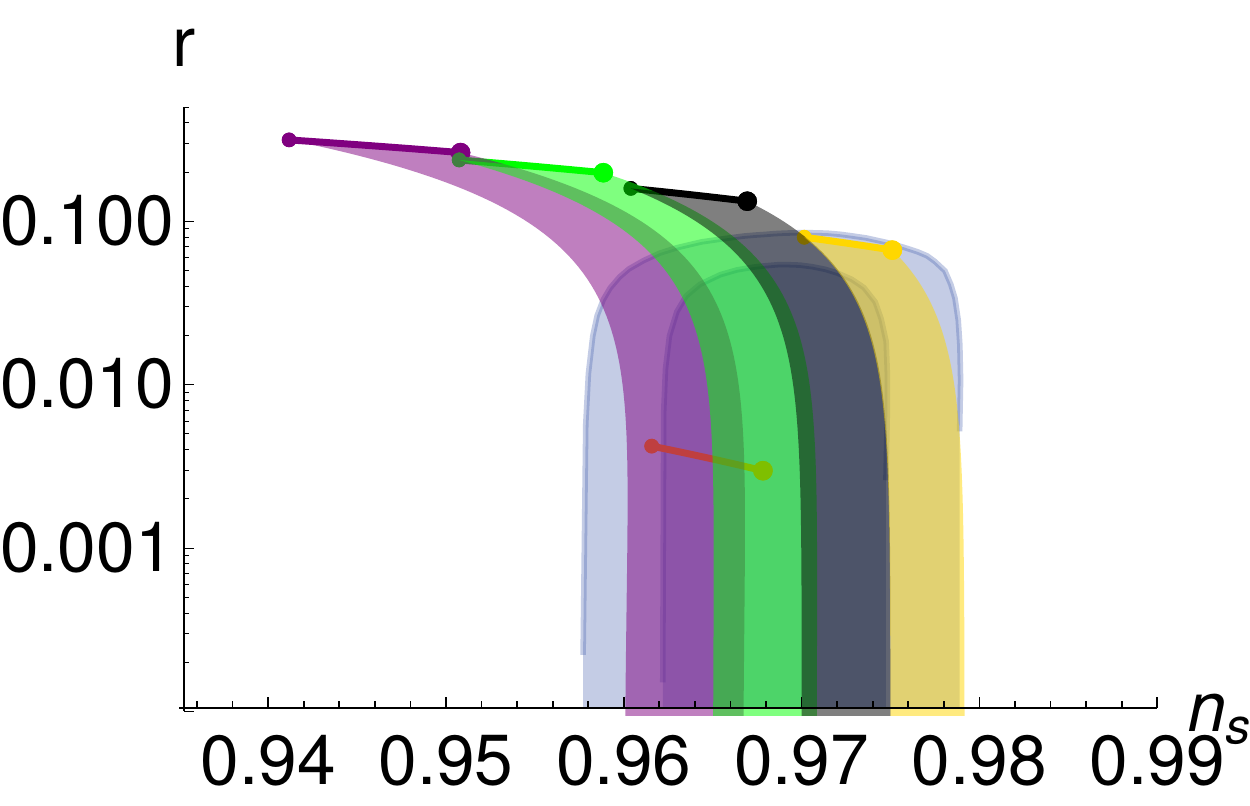}
\includegraphics[width=0.45 \textwidth]{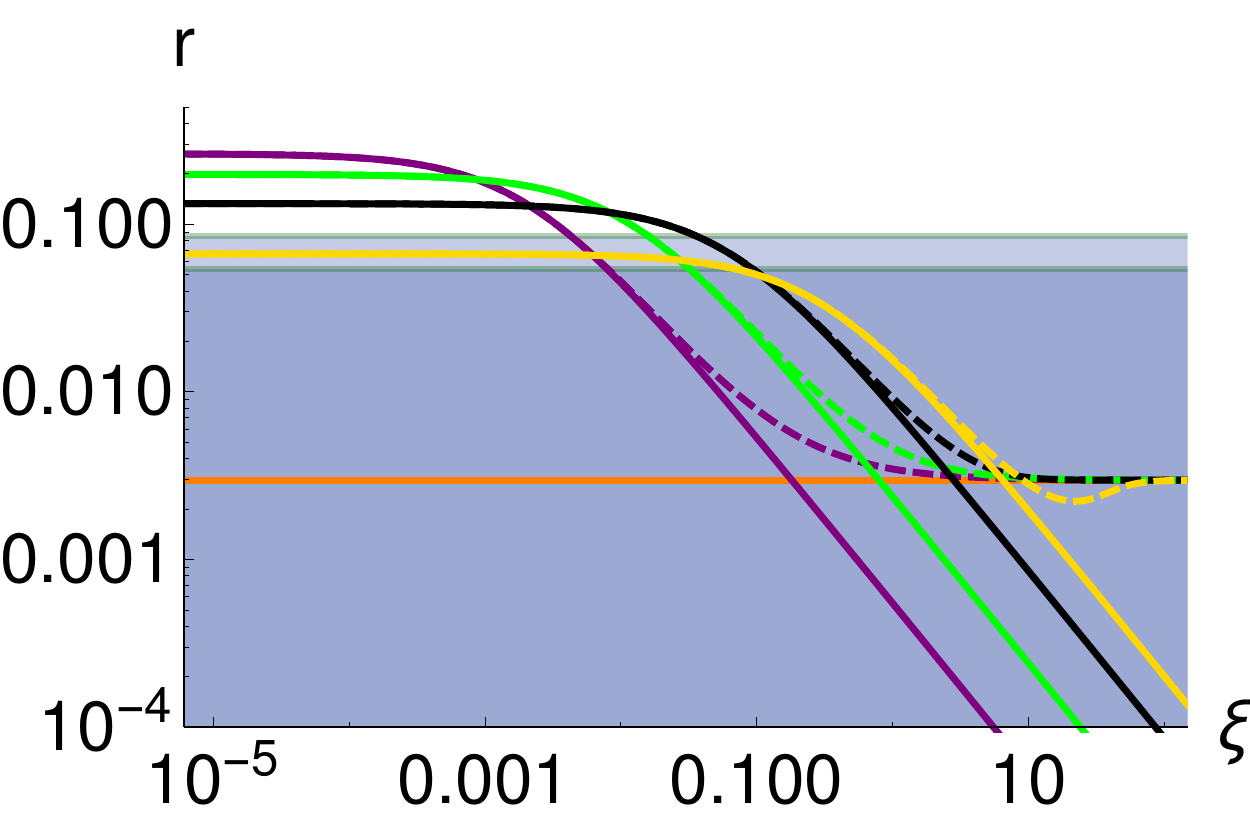}\\
    \caption{Tensor-to-scalar ratio $r$ as a function of $n_s$ with $N \in [50,60]$ $e$-folds in the Palatini formulation of the models described by Eq. (\ref{eq:f}) (upper panel); and $r$ as a function of $\xi$ for $N=60$ $e$-folds (lower panel, solid lines). Shown are predictions for $n=1/2$ (yellow), $n=1$ (black), $n=3/2$ (green) and $n=2$ (purple). For reference, we also plot the predictions of the corresponding $\alpha$-attractors (dashed) and Starobinsky inflation in metric gravity (orange).  The light blue areas present the 1, 2$\sigma$ constraints from the BICEP2/Keck data \cite{Array:2015xqh}. 
The data are extracted from figure 7 of Ref. \cite{Array:2015xqh}.
    }
   \label{Fig:r:vs:ns}
  \end{center}
\end{figure}

We can now provide a comparison between the metric and the Palatini formulation of the model described by Eqs. (\ref{nonminimal_action}), (\ref{eq:f}) and (\ref{eq:V}). The results are given in Fig. \ref{Fig:r:vs:ns}, where we plot $r$  as a function of $n_s$ with $N \in [50,60]$ $e$-folds 
(upper panel) and as a function of $\xi$ with $N=60$ $e$-folds (lower panel) for the Palatini formulation of the models described by Eq. (\ref{eq:f}). We considered $n=1/2$ (yellow), $n=1$ (black), $n=3/2$ (green) and $n=2$ (purple). For reference, we also plot the predictions of the corresponding $\alpha$-attractors (dashed) and Starobinsky inflation in metric gravity (orange). The light blue areas present the 1, 2$\sigma$ constraints from the BICEP2/Keck data \cite{Array:2015xqh}. 

We can see that due to the different field redefinition, Eq. (\ref{chi}), the Starobinsky attractor is lost in the Palatini gravity\footnote{It is however possible to recover the attractor behaviour in Palatini gravity with an {\it ad hoc} and less appealing choice of the functions $f(\phi)$ and $V(\phi)$. More details are illustrated in the following section.}. For $\xi\to\infty$, we find that
\bea
n_s &\simeq& 1 - \left( 1+\frac{n}{2} \right) \frac{1}{N} \,, \\ \nonumber
r &\simeq& 0 \, ,
\eea
which are clearly different from the usual Starobinsky solution, Eq. \eqref{starobinsky_nsr}.

In the future, the sensitivity of the CORE mission, $\sigma_r \simeq 4 \times 10^{-4}$ \cite{Remazeilles:2017szm}, will be enough to rule out (or confirm) Starobinsky inflation. Therefore, for a broad range of non-minimal coupling values $\xi$, our new scenario might be discriminated from the usual metric $\alpha$-attractors. 

\section{Attractors generated by Palatini gravity} \label{subsec:gen:Palatini}

In this last section, we present the Jordan frame functions $f(\phi)$ and $V(\phi)$ that correspond to the scalar potential (\ref{eq:metric:attractor}) in non-minimally coupled Palatini gravity, effectively recovering the attractor behavior. By using Eqs. (\ref{Omega}), (\ref{chi}) and (\ref{eq:U}), it is possible to show that this can be achieved with the following choices for $V(\phi)$ and $f(\phi)$:
\bea
 V(\phi) &=& \lambda f(\phi)^2 \Big( 1 - G(\phi) \Big) , \label{eq:V:P:alpha}\\
 f(\phi) &=& \frac{2}{3\alpha} \left( \frac{G(\phi)}{{\rm d} G(\phi)/{\rm d} \phi} \right)^2  \label{eq:f:P:alpha},
\eea
where $G(\phi)$ is an arbitrary differentiable function of $\phi$, $\lambda$ is a prefactor that can be adjusted to fit the normalization of Eq.\ (\ref{eq:metric:attractor}), and we have adopted the $\alpha$-attractor notation of Eq.\ (\ref{eq:alpha}). 
From Eq.\ (\ref{eq:f:P:alpha}) we can see that by choosing 
\be
G(\phi)=\left(\frac{\phi}{M_P}\right)^n, 
\ee
the Palatini attractors present the nice feature of having the non-minimal coupling always quadratic:
\be
f(\phi)=\frac{2}{3\alpha n^2} \phi^2 ,
\ee
which corresponds to the well-known case of induced-gravity inflation \cite{Spokoiny:1984bd,Accetta:1985du,Kaiser:1993bq,Kaiser:1994wj}. However, the corresponding scalar potential 
\be
V(\phi) =  \frac{4 \lambda}{9\alpha^2 n^4} \phi^4 \left[ 1 - \left(\frac{\phi}{M_P}\right)^n \right],
\ee
would not be renormalizable for $n>0$. The potential $V(\phi)$ could be renormalizable for certain $n<0$ choices but this case is not allowed since, in the attractor limit $\alpha \to 1$, it would lead to the forbidden case $f(\phi)<0$ (cfr.\ Eq.\ (\ref{chi}) or see \cite{Galante:2014ifa} and references therein). Moreover, choosing a suitable $G(\phi)$ so that $f(\phi) \approx 1 + c \, \phi^2$, would lead to an even more complicated $V(\phi)$.  Therefore, although inflationary attractors can be still obtained also in Palatini gravity, the price to pay is to consider more involved functions than in the metric case\footnote{It is also possible to consider a non-minimal kinetic term of the type $K(\phi) \partial_\mu \phi \partial^\mu \phi$ in addition to Eq.\ \eqref{nonminimal_action}, see e.g. Ref \cite{Koivisto:2005yc}. This will add even more freedom to reproduce the Einstein frame expression for the scalar potential in the case of $\alpha$-attractors. However, being beyond the purpose of the present article, we postpone such analysis to a forthcoming article.}.


\section{Conclusions}
\label{conclusions}

We performed an analysis of models of chaotic inflation where the inflaton field $\phi$ is coupled non-minimally to gravity, focusing on the {\it Palatini} theory. We showed that in this case the famous attractor property of simple chaotic inflation models is lost. Therefore, the existence of such property is not universal for models of inflation but depends on the underlying theory of gravity in a subtle way. The future CORE mission might be able to discriminate between the metric inflationary attractors and their Palatini counterparts. In light of these forthcoming data, it would be interesting to investigate universality of inflationary attractors also in other theories of gravity.

\section*{Acknowledgements}
We thank David I. Kaiser and Tomi Koivisto for useful correspondence. L.J. is supported by the Estonian Research Council grants IUT02-27 and PUT790, and by the ERDF Centre of Excellence project TK133. A.R. is supported by the Estonian Research Council grants IUT23-6, PUT1026, and by the ERDF Centre of Excellence project TK133. T.T. is supported by the U.K. Science and Technology Facilities Council grant ST/J001546/1.

\bibliography{palatini_models}

\end{document}